\documentclass[pdftex,twocolumn,superscriptaddress]{revtex4}

\usepackage{color,graphicx}           
\usepackage{amssymb}                  
\usepackage{lineno}                   

\begin{document}

\title{The role of static disorder in negative thermal expansion in ReO$_3$}

\author{Efrain E. Rodriguez}
\affiliation{Los Alamos National Laboratory, Lujan Neutron
Scattering Center, MS H805, Los Alamos, NM 87545, USA}
\affiliation{Materials Department and Materials Research Laboratory,
University of California, Santa Barbara, CA 93106, USA}
\author{Anna Llobet}
\affiliation{Los Alamos National Laboratory, Lujan Neutron
Scattering Center, MS H805, Los Alamos, NM 87545, USA}
\author{Thomas Proffen}
\affiliation{Los Alamos National Laboratory, Lujan Neutron
Scattering Center, MS H805, Los Alamos, NM 87545, USA}
\author{Brent C. Melot}
\author{Ram Seshadri}
\affiliation{Materials Department and Materials Research Laboratory,
University of California, Santa Barbara, CA 93106, USA}
\author{Peter B. Littlewood}
\affiliation{Cavendish Laboratory, University of Cambridge,
Madingley Road, Cambridge, CB3 OHE, UK}
\author{Anthony K. Cheetham}
\email{akc30@cam.uc.uk}
\affiliation{Department of Materials Science
and Metallurgy, University of Cambridge, Pembroke Street, Cambridge,
CB2 3QZ, UK}

%
%
\begin{abstract} Time-of-flight neutron powder diffraction and
specific heat measurements were used to study the nature of thermal
expansion in rhenium trioxide, an electrically conducting oxide with
cubic symmetry.  The temperature evolution of the lattice parameters
show that ReO$_3$ can exhibit negative thermal expansion at low
temperatures and that the transition from negative to positive
thermal expansion depends on sample preparation; the single crystal
sample demonstrated the highest transition temperature, 300 K, and
largest negative value for the coefficient of thermal expansion,
$\alpha = -1.1(1)\times10^{-6}$ K $^{-1}$. For the oxygen atoms, the
atomic displacement parameters are strongly anisotropic even at 15
K, indicative of a large contribution of static disorder to the
displacement parameters.  Further inspection of the temperature
evolution of the oxygen displacement parameters for different
samples reveals that the static disorder contribution is greater for
the samples with diminished NTE behavior. In addition, specific heat
measurements show that ReO$_3$ lacks the low energy Einstein-type
modes seen in other negative thermal expansion oxides such as
ZrW$_2$O$_8$.
\end{abstract}

\maketitle
%

\section{Introduction}
Metals that contract upon heating are exceedingly rare, if they
exist at all. This observation can be understood by considering the
types of cohesive interactions bonding atoms into solids.  As a
material is heated, interatomic distances expand due to the
asymmetric shape of the potential energy well and therefore to
anharmonic effects.\cite{Kittel,sleight_1998b} A chemical bond with
a deeper potential well will confine atomic vibrations to remain
more harmonic and consequently, covalently bonded materials such as
diamond exhibit lower thermal expansion than metals such as copper.
In some covalent crystals with open framework structures, the low
thermal expansion can even become negative.  Among the negative
thermal expansion (NTE) materials most studied are framework oxides
such as ZrW$_2$O$_8$\cite{mary_1996,evans_1997,ernst_1998} and more
recently metal cyanide frameworks such as
Zn(CN)$_2$,\cite{goodwin_2005,chapman_2006b} all of which are
electrical insulators or semiconductors.  In contrast, the oxide
ReO$_3$ is a superb electrical conductor with a room temperature
resistivity close to that of copper.\cite{ferretti_1965,king_1971}
Recently, Chatterji \textit{et al.} have demonstrated through
neutron powder diffraction and first-principles calculations that
ReO$_3$ also exhibits isotropic NTE,\cite{chatterji_2008} making it
the only known simple binary oxide to combine metallic conductivity
with NTE.

\begin{figure} [!b] \centering
\includegraphics[angle=0.0]{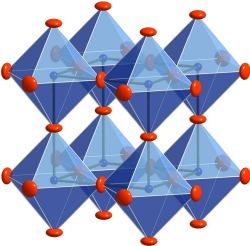}
\caption {The crystal structure of the cubic perovskite ReO$_3$.
ReO$_6$ octahedra connect via corner-sharing in infinite chains in
all three directions. The atomic displacement ellipsoids of the
oxygen atoms (in red) are represented at 90 \% probability; their
flat shape indicates transverse displacement normal to the Re--O--Re
bonds and into the void in the middle of the cubic
cell.}\label{ReO3_structure}
\end{figure}
A topological feature that can lead to NTE behavior in an open
framework structure is a linear chain consisting of a ligand
coordinated to two metals or \textit{vice
versa}.\cite{sleight_1998b,sleight_1998} For example, simple binary
oxides such as Cu$_2$O, Ag$_2$O (O--M--O
chains)\cite{mittal_2007}and certain phases of SiO$_2$ (Si--O--Si
chains)\cite{attfield_1998b,lightfoot_2001} demonstrate NTE
behavior. While this is not a requisite for NTE behavior
(magnetostriction and phase transitions can also result in NTE), it
is a simple principle by which new NTE materials can be pursued. The
structure of rhenium trioxide contains such linear chains in a three
dimensional framework consisting of corner-sharing ReO$_6$ octahedra
(See Fig.\,\ref{ReO3_structure}). The simple ReO$_3$ structure is a
prototype for closely related materials such as the perovskite
ferroelectrics, ferromagnets, and high T$_c$ superconductors and can
also be thought of as the cubic perovskite ABO$_3$ with a void in
the place of the $A$ cation.

Thermal expansion investigations of ReO$_3$ have presented
conflicting results.  A laser interferometry study on a single
crystal concluded that ReO$_3$ displayed NTE between 100\,K and
340\,K,\cite{matsuno_1978}. In contrast to the single crystal work,
a catalogue of powder X-ray diffraction (XRD) studies reported the
coefficient of thermal expansion of ReO$_3$ to be small but positive
for all temperature ranges.\cite{taylor_1985} The latter study has
been more frequently cited in reviews of NTE in oxide
frameworks,\cite{sleight_1998b,tao_2003b} leaving ReO$_3$ largely
unrecognized as a potential NTE material until the recent article by
Chatterji \textit{et al}, which showed through diffraction studies
NTE up to 200 K.\cite{chatterji_2008} In the present study, we show
that ReO$_3$ can exhibit isotropic NTE up to room temperature, that
the positive to negative transition is dependent on sample quality,
and that specific heat measurements show a lack of any low energy
Einstein-type modes that lead to the NTE behavior in other oxides.
Furthermore, we show through careful analysis of the atomic
displacement parameters the role of static disorder of the oxygen
atoms in impeding NTE behavior and therefore explain the discrepancy
between past thermal expansion studies of ReO$_3$.

%
\section{Experimental Methods}
The thermal expansion properties of various ReO$_3$ powder samples
were investigated by performing variable temperature neutron powder
diffraction (NPD) with time-of-flight neutron data.  The first
sample investigated, labeled ReO$_3$-a throughout this paper, was a
powder prepared by the decomposition of an Re$_2$O$_7$--1,4 dioxane
adduct at 140\,$^{\circ}$C as first described by Nechamkin
\textit{et al}.\cite{nechamkin_1951}  The second sample studied,
ReO$_3$-b, was a powder purchased from Alfa Aesar (99.99\% chemical
purity). Finally, the third sample, ReO$_3$-c, was prepared by
crushing single crystals of ReO$_3$ grown by chemical vapor
transport (CVT). By using HgCl$_2$ as the transport agent in an
evacuated Pyrex glass ampoule with ReO$_3$ powder (ReO$_3$-b as the
starting powder), the crystals were grown across a temperature
gradient of about 100\,$^{\circ}$C as described by Feller \textit{et
al.}\cite{feller_1998}. In addition to growing single crystals, CVT
also separates the lower oxide phase impurities from the freshly
grown ReO$_3$ crystals. Hence, sample ReO$_3$-c was expected to be
of higher chemical and phase purity than the others.

NPD patterns were obtained on the high intensity powder
diffractometer (HIPD) at the Lujan Neutron Scattering Center at the
Los Alamos Neutron Science Center (LANSCE). HIPD is suitable for
studying the behavior of materials as a function of temperature and
pressure due to its relatively high data acquisition rates.  HIPD
data were collected on all samples between 15\,K and 300\,K.  The
Lujan Center employs a pulsed spallation neutron source, so the NPD
histograms were collected in time-of-flight mode on fixed detector
banks.  The backscattering detector banks, located $\pm153 ^{\circ}$
with respect to the incident beam, provide the highest resolution
pattern and higher-index reflections while the normal ($\pm90
^{\circ}$) and forward scattering ($\pm40 ^{\circ}$) banks have a
lower resolution but probe higher $d$-spacings.  In the study by
Chatterji \textit{et al.}, the neutron diffraction measurements were
performed on a constant wavelength source ($\lambda = 1.359\pm0.001$
{\AA}),\cite{chatterji_2008} limiting the diffractogram up to $8.5$
{\AA}$^{-1}$ in momentum transfer $Q=4\pi\sin\theta/\lambda$. More
accurate structural parameters such as atomic displacement
parameters can be accessed with time-of-flight neutron data since
higher $Q$ ranges can be probed. In our measurements, a $Q_{max}$ of
$16$ {\AA}$^{-1}$ was used for all the structural refinements.

The specific heat data were collected upon warming from 2\,K to
20\,K on a 28.5 mg single crystal using a quasiadiabatic method as
implemented on a Quantum Design Physical Property Measurement System
(PPMS).  Like sample ReO$_3$-c, the single crystal was grown by CVT.

%
\section{Results and Discussion}
\begin{figure} [!b] \centering
\includegraphics{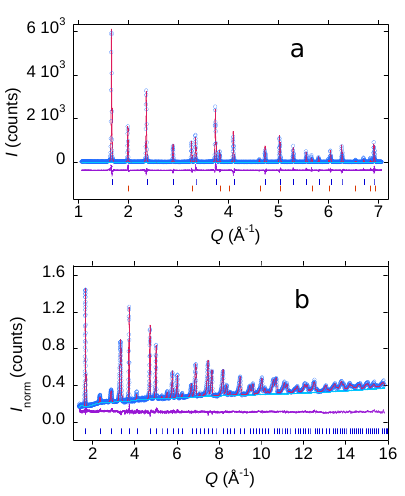}
\caption{(a) Observed profile of ReO$_3$ from X-rays shown in open
circles and the calculated profile as a smooth red line.  The X-ray
powder patterns contain both sample ReO$_3$-c and a Si standard.
Upper tickmarks correspond to ReO3-c and lower to the Si standard.
(b) Observed profile from neutrons is shown in open circles and the
calculated as a smooth red line for sample ReO$_3$-c only with
corresponding Bragg reflection tickmarks below.  The difference
curve between the observed and calculated patterns are represented
by the smooth purple line below the powder profiles.}
\label{ReO3_xrd_hipd}
\end{figure}

\begin{figure} [!b] \centering
\includegraphics[angle=0.0]{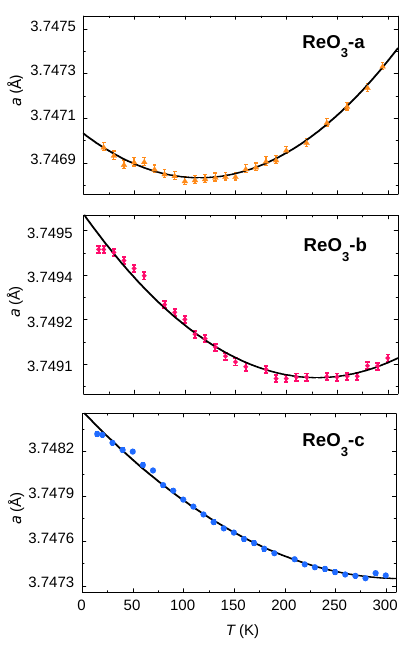}
\caption{Temperature evolution of the lattice parameters obtained
from neutron diffraction for samples ReO$_3$-a, ReO$_3$-b, and
ReO$_3$-c. The smooth black curves are the least-squares, polynomial
fits to the lattice parameters.  Error bars are obtained from
standard uncertainties in the structural
refinements.}\label{lattice_temp}
\end{figure}

Least squares structural refinements by the Rietveld method were
performed with all six histograms of the NPD data using the GSAS
software package.\cite{gsas} In space group $Pm\overline{3}m$, only
four variables were refined for each temperature: the lattice
parameter; the isotropic displacement parameter $U_{iso}$ of Re, and
the anisotropic displacement parameters $U_{11}$ ($=U_{22}$), and
${U_{33}}$ of O.  To determine the lattice parameters of the sample
accurately, X-ray powder diffraction was performed at room
temperature with Si Standard Reference Material 640c purchased from
National Institute of Standards (See Fig.\,\ref{ReO3_xrd_hipd}a).
The neutron powder profile for the sample prepared from single
crystals, ReO$_3$-c, is shown in Fig.\,\ref{ReO3_xrd_hipd}b.  In
this sample, the lattice parameter plotted as a function of
temperature clearly demonstrates NTE behavior from 15\,K up to
300\,K (See Fig.\,\ref{lattice_temp}c).  NTE in the other powder
samples was indeed observed but limited to a narrower temperature
range (See Fig.\,\ref{lattice_temp}a,b). For sample ReO$_3$-a, the
NTE behavior was observed only up to 110\,K, and for ReO$_3$-b, only
up to 220\,K.  Clearly sample quality has an effect on the thermal
expansion properties and this may, in part, explain the discrepancy
between past findings on the thermal expansion behavior of ReO$_3$
and the reason why Chatterji \textit{et al.} found NTE only up to
200 K from their neutron measurements while their dynamical lattice
calculations predicted NTE up to 350\,K.\cite{chatterji_2008}

The linear coefficient of thermal expansion for a cubic system
$\alpha$ is defined as $\alpha = (1/a)\times\partial{a}/\partial{T}$
where $a$ is the cubic cell parameter and $T$ the temperature.  A
least squares polynomial fit (second-order) to the lattice
parameters obtained from the neutron data affords $a(T)$, which can
be differentiated with respect to $T$ and divided by the
experimental values of $a$ to obtain $\alpha$ as a function of $T$.
For sample ReO$_3$-c the average $\alpha$ below 300\,K was found to
be $-1.1(1)\times10^{-6}$\,K$^{-1}$.  For sample ReO$_3$-a, $\alpha$
was found to be $-3.5(6)\times10^{-7}$\,K$^{-1}$, and
$-6.4(5)\times10^{-7}$\,K$^{-1}$ for sample ReO$_3$-b.  Thus,
lowering of the transition temperature $T_{NTE\rightarrow PTE}$ is
correlated with a smaller, negative value for $\alpha$.

For sample ReO$_3$-c, the values for $\alpha(T)$ are close to those
determined by Matsuno \textit{et al.} in their interferometry study
on single crystals where the value of $\alpha$ is
$-2\times10^{-6}$\,K$^{-1}$ at about 120\,K and slowly rises to
$-1\times10^{-6}$ K$^{-1}$ at about 230\,K.\cite{matsuno_1978} While
we did not make measurements above room temperature, Matsuno
\textit{et al.} found $\alpha$ to be positive and smaller than
$2\times10^{-6}$\,K$^{-1}$ between 340\,K and
500\,K.\cite{matsuno_1978} In an XRD study carried out on single
crystals as well as powder samples, Chang \textit{et al.} found a
value of $1.1(1)\times10^{-6}$ K$^{-1}$ above room
temperature.\cite{chang_1978} The volumetric $\alpha$ calculated by
Chatterji \textit{et al.} is not linear as in our simplified model
from the fitted curves, but nevertheless shows a volumetric $\alpha$
($=1/3$ linear $\alpha$) varying between $-2\times10^{-6}$ to zero
below 350 K.\cite{chatterji_2008}

\begin{figure} [!b] \centering
\includegraphics[angle=0.0]{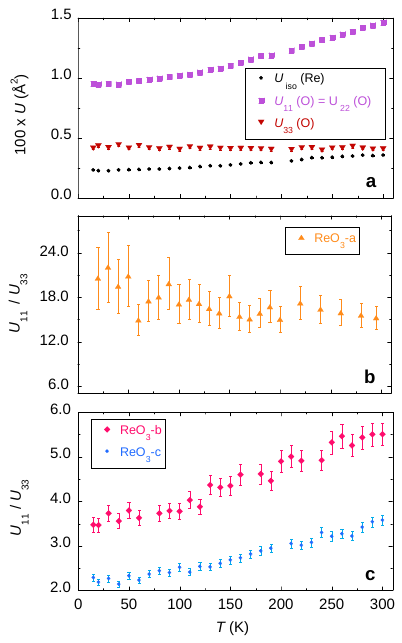}
\caption{(a) Temperature evolution of the atomic displacement
parameters for ReO$_3$-c, where $U_{iso}$ represents the rhenium
isotropic displacement parameters and $U_{11},U_{22},U_{33}$ the
oxygen anisotropic displacement parameters. (b) The ratio
$U_{11}$/$U_{33}$ for sample ReO$_3$-a and (c) the same ratio for
samples ReO$_3$-b and ReO$_3$-c.  $U_{11}$ and $U_{22}$ represent
atomic displacement of the oxygen atoms normal to the Re--O--Re
bond, while $U_{33}$ to displacement along the bond.  Error bars,
shown within the symbols in (a), are obtained from the standard
uncertainties in the structural refinements.}\label{thermal_temp}
\end{figure}

The atomic displacement parameters obtained from the NPD data offer
some clues as to the mechanism responsible for NTE in ReO$_3$. Due
to the high range of momentum transfer $Q$ available from
time-of-flight neutrons as well as the lack of any decay with $Q$ in
the atomic form factors for neutron diffraction, accurate atomic
displacement parameters were obtained from the Rietveld analysis
(See Fig.\,\ref{thermal_temp}a). The Re atom is located on the
special position $1a$ (site symmetry $\overline{3}m$), which
constrains the atomic displacement parameter to remain isotropic.
The oxygen atoms are located on the position $3c$ (site symmetry
$4/mmm$), constraining them to have only two values, $U_{11}$ ($=
U_{22}$) and $U_{33}$. While the displacement of the O atoms along
the Re--O--Re direction remains fairly constant, the transverse
motion increases upon heating, as also observed by Chatterji
\textit{et al}.\cite{chatterji_2008}  The most striking feature is
the gap between $U_{11}$ ($=U_{22}$) and $U_{33}$ as represented by
the flatness of the displacement ellipsoids of the O atoms (See
Fig.\,\ref{ReO3_structure}). The anisotropy only increases with
temperature, suggesting that the transverse motion contracts the
cubic lattice upon heating and is therefore the most likely
mechanism for the observed NTE behavior.

Further inspection of the oyxgen displacement parameters of all
three samples reveals differences that may explain the varying
thermal expansion behavior.  Though sometimes loosely referred to as
thermal parameters or Debye-Waller factors, the values for $U_{ij}$
have contributions from both dynamic and static displacements of the
atoms.  Extrapolation of the $U_{ij}$ parameters for ReO$_3$ down to
0\,K reveals non-zero values, which indicates significant
contribution from static displacement, especially in the case of
$U_{11}$ (See Fig.\,\ref{thermal_temp}a).  The large gap between
$U_{11}$ and $U_{33}$ at low temperatures was observed for all three
samples.  A useful parameter to describe the extent of anisotropy in
the atomic displacements is the ratio of the maximum value to the
minimum value of $U_{ij}$;\cite{trueblood_1996}  in our case, it is
$U_{11}$/$U_{33}$.  When plotted versus temperature (See
Fig.\,\ref{thermal_temp}b,c), this measure of anisotropy reveals
that it is constant (within the error bars) and large for sample
ReO$_3$-a, while it is increases upon heating for ReO$_3$-b and
ReO$_3$-c. Most importantly, the higher $U_{11}$/$U_{33}$ ratios for
ReO$_3$-a and ReO$_3$-b than for ReO$_3$-c at low temperatures
suggest more static disorder in the first two, which should affect
the lattice dynamical properties.  The $M3$ mode calculated by
Chatterji \textit{et al} involves rotation of the ReO$_6$ octahedra
in the $a-b$ plane and arises from transverse motion of the O atoms
in this plane.\cite{chatterji_2008}  Since the $M3$ mode has the
largest contribution to the negative value for $\alpha$ at low
temperatures, static displacement of the O atoms, transversely from
the Re--O--Re bond, would diminish the NTE behavior.

\begin{figure} [!t] \centering
\includegraphics[angle=0.0]{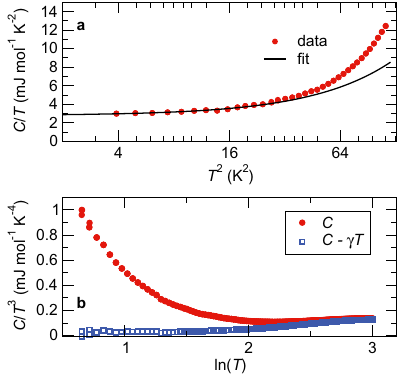}
\caption{Specific heat data for a single crystal ReO$_3$. (a) Data
fit below 4\,K  to the function $C/T = \gamma + \beta T^2$ (where
$\Theta^3 = 12/5\pi^4 R /\beta$; $R$ is the gas constant). The
fitting allows the Sommerfeld coefficient $\gamma$ =
3.6\,mJ\,mol$^{-1}$\,K$^{-2}$ and a Debye temperature of $\Theta$ =
344\,K to be obtained. (b) Data plotted as $C/T^3$ \textit{vs.} $\ln
T$ in order to reveal any low energy phonon spectral weight. The
absence of any peaks indicates a lack of contribution from any
Einstein-type modes; the Debye contribution appears as a constant up
to $\Theta$ and the rise at low $\ln T$ is due to the electronic
contribution ($\gamma T$) of the delocalized
electrons.}\label{heat_cap}
\end{figure}
Specific heat measurements on a single crystal of ReO$_3$ were
carried out to understand the type of lattice vibrations
contributing to the specific heat.  The low temperature behavior in
the heat capacity of ReO$_3$ is that of a straightforward Debye
solid, with a Debye temperature $\Theta$ of 344\,K and a Sommerfeld
coefficient $\gamma$ of 2.8\,mJ\,mol$^{-1}$\,K$^{-2}$ (See
Fig.\,\ref{heat_cap}a).  These values agree very well with the
values of 327\,K for $\Theta$ and 2.85\,mJ\,mol$^{-1}$\,K$^{-2}$ for
$\gamma$ found in the early studies by King \textit{et
al.}\cite{king_1971}  Fig.\,\ref{heat_cap}b shows the specific heat
over $T^3$ vs. $\ln T$, which is an approximate representation of
the one-dimensional phonon density of states of a
solid.\cite{ramirez_1998,junod_1983}  A signature of an Einstein
mode in this representation is a Gaussian-like peak--clearly missing
in the low temperature region of the specific heat.  These
measurements strengthen the argument by Chatterji \textit{et al.}
from their lattice dynamical calculations that NTE in ReO$_3$ is
caused by low energy acoustic modes with the $M3$ mode contributing
the most to the NTE behavior.\cite{chatterji_2008} Also, the
specific heat results rule out the suggestion by Matsuno \textit{et
al.} that a low energy, bending-like optical mode is responsible for
NTE in ReO$_3$.\cite{matsuno_1978}

Although both ZrW$_2$O$_8$ and ReO$_3$ have cubic symmetry, it is
interesting to compare how their structures lead do different NTE
behavior.  The primitive cubic structure of ReO$_3$ (given the
absence of the $A$ cation in its structure and thus an extremal
value of the perovskite tolerance factor) is of course the
consequence of metallicity: the large volume is maintained by the
Fermi pressure of the delocalized d-electrons in the $\pi^*$
conduction band. Nevertheless, the proximity to a `buckling
transition' is revealed by a pressure-induced collapse of the
structure produced by cooperative counter-rotation of the
octahedra,\cite{jorgensen_2000,batlogg_1984} in this case very
similar to ZrW$_2$O$_8$.\cite{jorgensen_1999,pantea_2006} The linear
$\alpha$ obtained for ReO$_3$ is smaller than that observed in
ZrW$_2$O$_8$, where $\alpha=-8.7\times10^{-6}$ K$^{-1}$.  Indeed,
the smaller absolute value of the NTE and the smaller temperature
range at which it occurs in ReO$_3$ compared to ZrW$_2$O$_8$ is
consistent with the much weaker precursor effects in the
pressure-induced collapse in the former than in the
latter.\cite{pantea_2006}  In turn, the weakness of the fluctuations
in ReO$_3$ is possibly a consequence of it possessing fewer soft
degrees of freedom than the more open framework structure of
ZrW$_2$O$_8$.  In addition, the low energy Einstein-type modes
observed in ZrW$_2$O$_8$ and implicated in that material's NTE
behavior,\cite{ernst_1998,ramirez_1998} are clearly missing in
ReO$_3$.

The relatively small NTE effect observed in ReO$_3$ remains,
nevertheless, remarkable because it is isotropic and unprecedented
for a material with the electronic properties of a simple metal.
While classifying an oxide as a simple metal is atypical, the
free-electron model works well enough in ReO$_3$ that it qualifies
under the scheme of electrical transport properties (other examples
include the Na$_x$WO$_3$ bronzes and RuO$_2$).\cite{cox} Most simple
metals have a much larger $\alpha$ than that of ReO$_3$. For
example, for copper, $\alpha$ is $16.5\times10^{-6}$\,K$^{-1}$ at
room temperature.\cite{crc_handbook} Another notable exception is
the intermetallic YGaGe, which has such small $\alpha$ that its
behavior has been termed zero thermal expansion.\cite{salvador_2003}
Like YGaGe, ReO$_3$ is also remarkable in that its $\alpha$ above
room temperature is as low as in a highly covalent material like
diamond where $\alpha$ is
$1\times10^{-6}$\,K$^{-1}$.\cite{chang_1978} While ReO$_3$ does
exhibit highly metallic behavior, the strong covalency between the
Re d-states and O p-states and its unique framework topology allow
for negative thermal expansion, demonstrating the propensity for
transition metal oxides to challenge our conventions on bonding and
properties in solid materials.  Furthermore, ReO$_3$'s relatively
simple structure has provided a way to quantify the role of static
disorder of the oxygen atoms on interesting behavior such as
negative thermal expansion.
%

\section{Acknowledgements}
This work has benefited from the use of HIPD at the Lujan Center at
Los Alamos Neutron Science Center, funded by DOE Office of Basic
Energy Sciences. Los Alamos National Laboratory is operated by Los
Alamos National Security LLC under DOE Contract DE-AC52-06NA25396.
We would also like to thank J. C. Lashley and A. Lawson from Los
Alamos for stimulating discussions and L. L. Daeman, also from Los
Alamos, for help with sample preparation.  BCM and RS acknowledge
the National Science Foundation for support through a Career Award
(NSF-DMR-0449354), and for the use of MRSEC facilities (Award
NSF-DMR0520415).
%

%
\end{document}